\documentstyle[preprint,prc,eqsecnum,aps,epsf]{revtex}

\tightenlines   

\begin{document}
\draft

\title{Model Study of Three-Body Forces in the Three-Body Bound
State}

\author{H. Liu, Ch. Elster}
\address{
Institute of Nuclear and Particle Physics,  and
Department of Physics, \\ Ohio University, Athens, OH 45701}

\author{ W. Gl\"ockle}
\address{
 Institute for Theoretical Physics II, Ruhr-University Bochum,
D-44780 Bochum, Germany.}

\vspace{10mm}

\date{\today}

\maketitle

\begin{abstract}
The Faddeev equation for the three-body bound state 
with two- and three-body forces is solved directly as three-dimensional integral equation.
The numerical feasibility and stability of the 
algorithm, which does not employ partial wave decomposition is
demonstrated. The
three-body binding energy and the full wave function are calculated with
Malfliet-Tjon-type two-body potentials and scalar two-meson exchange 
three-body forces. 
For two and three body forces of ranges and strengths typical of nuclear forces the single 
particle momentum distribution and the two-body correlation function are similar  to the ones found
for realistic nuclear forces.

\end{abstract}

\vspace{10mm}


\pagebreak




\section{Introduction}

Three-body Faddeev equations for the bound state using local forces directly
without finite rank expansion have been solved since the pioneering work by 
Malfliet and Tjon \cite{malfliet69-1} and Osborn \cite{osborn}.
After this, a huge body of work by different groups followed, and 
calculations using the Faddeev equations were
either performed in momentum space, see e.g. 
\cite{pickle,stadler,stadler2,nbench}, 
in configuration space \cite {lavern,payne,schell0}, or in a hybrid fashion using
both spaces \cite{wu}.
However, the technique was always based on an angular momentum decomposition,
which including spin and isospin degrees of freedom lead to quite a large set of states. 
For instance allowing NN forces to act in all states of total angular momenta $j$ up to
$j$=6, which
is necessary to control the $^3$H binding energy within 2 keV require 102 angular momentum
and isospin combinations (in the literature often called channels). 
In view of this very
large number of interfering terms and having in mind that computer resources are  
still increasing steadily, it 
appears natural to give up such an expansion and  work directly with momentum
vectors or position vectors in configuration  space as it is common in
Greens Function Monte Carlo methods \cite{gfmc}. 
We started an approach based on momentum vectors as variables for the 
three-boson bound state \cite{elster98-1},  which then could be extended to the
three-boson continuum in a straightforward fashion \cite{scatt1}.
In configuration space Faddeev equations have been solved applying vector variables 
for pure Coulomb bound-state problems, namely the $e^-e^-e^+$ and the
$pp\mu^-$ systems \cite{kvin}. 
Including spin  and isospin degrees of freedom together with realistic nucleon-nucleon 
(NN) forces,
a scheme based on momentum vectors has been applied to 
NN scattering leading to integral
equations in two variables \cite{imam1}, which can easily be solved on present day computers.

Angular momentum decompositions even for three bosons require  quite a tedious algebra
and care in a computational implementation \cite{gloeckle-book}
due to often nontrivial cancellations between
partial wave terms.
This is especially true if one includes 
three-body forces \cite{coon,hueber97-1}. 
In contrast, the formulation in terms of momentum vectors of the
one Faddeev equation for identical particles is very straightforward and transparent. 
In this paper we extend the approach from Ref.~\cite{elster98-1}
and include three-body forces for an investigation of the three-body bound state.
This study is meant only to demonstrate the ease in solving the Faddeev equation including three-body
forces, and working directly with momentum vector variables. Since we do not yet include spin and
isospin degrees of freedom we do not address physical questions directly related to the physical three
nucleon system, but the numerical examples given can shed some light on the wave function properties
of three nucleons.   

The paper is organized as follows. In Section II the Faddeev equation including two- and
three-body forces is formulated in terms of momentum vectors, and its solution,
especially the
intermediate integrations are layed out in detail. Section III displays our choice of
forces and provides the necessary numerical insight for achieving an accurate solution.
Wave function properties are displayed in Section IV.
Finally we summarize in Section V and provide an outlook.

\section{Three-Body Bound State Equation with Three-Body Force}

The bound state of three identical particles which interact via pairwise
forces $V^i=V_{jk}$ ($i,j,k=1,2,3$ and cyclic permutations thereof) and
a genuine three-body force $V_{123}$ is given by the Schr\"odinger
equation which reads in integral form
\begin{equation}
|\Psi\rangle =
G_{0}(\sum^{3}_{i=1}V^{i}+V_{123})|\Psi\rangle.
\label{eq:2.1}
\end{equation}
Here the free propagator is given by
 $G_{0}=(E-H_{0})^{-1}$, where $H_0$ stands for the free Hamiltonian
and $E$ for the binding energy of the three-body system. Any
three-body force $V_{123}$ 
can be decomposed into three different pieces as
\begin{eqnarray}
V_{123}=\sum^{3}_{i=1}V^{(i)}_{4},
\label{eq:2.2}
\end{eqnarray}
such that $V^{(i)}_{4}$ is symmetric under the exchange of particles
$j$ and $k$ ($j \neq i \neq k$). The decomposition suggested in
Eq.~(\ref{eq:2.2}) is natural for e.g. realistic $\pi\pi$ 3N forces,
which are considered at present in all available 3N forces.
Introducing  Faddeev components
$|\Psi\rangle=\sum\nolimits^{3}_{i=1}|\psi_{i}\rangle$ with
\begin{equation}
|\psi_{i}\rangle = G_{0}(V_{i}+V^{(i)}_{4})|\Psi\rangle 
\label{eq:2.3}
\end{equation}
leads to three coupled integral equations
\begin{eqnarray}
|\psi_{i}\rangle&=&G_{0}t_{i}\sum_{j\neq i}
|\psi_{j}\rangle+(1+G_{0}t_{i})G_{0}V^{(i)}_{4}\sum_{j}
|\psi_{j}\rangle \nonumber \\
&=&G_{0}\{t_{i}\sum_{j\neq i}
|\psi_{j}\rangle+(1+t_{i}G_{0})V^{(i)}_{4}\sum_{j}
|\psi_{j}\rangle\}.
\label{eq:2.4}
\end{eqnarray}
The operator $t_{i}$ describes the two-body $t$-matrix in the
subsystem $jk$. If we consider identical particles (here bosons,
since we are omitting spin), the three-nucleon wave function
$|\Psi\rangle$ has to be totally symmetric. As a consequence, 
the Faddeev components
$|\psi_{i}\rangle$ are
identical in their functional form, only the particles are
permuted. Thus it is sufficient to consider only one component, e.g. 
\begin{equation}
|\psi_1\rangle=G_{0}t_1P|\psi_1\rangle +(1+G_{0}t)V^{(1)}_{4}(1+P)
|\psi_1\rangle
\label{eq:2.5}
\end{equation}
In the following the index (1) will be dropped.
The permutation operator $P$ is given as
$P=P_{12}P_{23}+P_{13}P_{23}$
and the total wave function reads 
\begin{eqnarray}
|\Psi\rangle=(1+P)|\psi\rangle.
\label{eq:2.6}
\end{eqnarray}
In order to solve  Eq.~(\ref{eq:2.5}) standard Jacobi
momenta are used,
\begin{eqnarray}
{\mathbf{p}}_{i}&=&
\frac{1}{2}\left({\mathbf{k}}_{j}-{\mathbf{k}}_{k} \right) \nonumber \\
{\mathbf{q}}_{i}&=&
\frac{2}{3}\left({\mathbf{k}}_{i}-
\frac{1}{2}\left({\mathbf{k}}_{j}+{\mathbf{k}}_{k}
\right) \right),
\label{eq:2.7}
\end{eqnarray}
where $ijk=123$ and cyclic permutations thereof. For later clarification
we  label the
coordinates with $ijk=123$ as system of `type ($1$)', the ones with 
$ijk=231$ as `type ($2$)' and the ones with $ijk=312$ as `type ($3$)'. 
With the Jacobi momenta from Eq.~(\ref{eq:2.7}) and omitting 
the arbitrarily chosen index 1, Eq.~(\ref{eq:2.5}) reads
\begin{eqnarray}
\langle {\mathbf{pq}}|\psi\rangle=\frac{1}{E-\frac{p^{2}}{m}-\frac{3q^{2}}{4m}}
\langle {\mathbf{pq}}|tP + V_{4}(1+P)+tG_{0}V_{4}(1+P)|\psi\rangle.
\label{eq:2.8}
\end{eqnarray}
Introducing the symmetrized two-nucleon $t$-matrix
\begin{eqnarray}
t_{s}({\mathbf{p}},{\mathbf{q}};E)=
t({\mathbf{p}},{\mathbf{q}};E)+t(-{\mathbf{p}},{\mathbf{q}};E)
\label{eq:2.9}
\end{eqnarray}
and explicitly working out the permutation operator $P$ in the 
first term of Eq.~(\ref{eq:2.8}) leads to
\begin{eqnarray}
\langle {\mathbf{pq}}|\psi\rangle&=&\frac{1}{E-\frac{p^{2}}{m}-\frac{3q^{2}}{4m}}
\Biggl [ 
\int d^{3}q' t_{s}\left( {\mathbf{p}},\frac{1}{2}{\mathbf{q}}+{\mathbf{q}}';E-\frac{3}{4m}q^{2} \right)
\langle {\mathbf{q}}+\frac{1}{2}{\mathbf{q}}',{\bf q'}|\psi\rangle  \nonumber \\
&+&
\langle {\mathbf{p}}{\mathbf{q}}|V_{4}(1+P)|\psi\rangle
+\frac{1}{2}\int d^{3} \tilde{p}
\frac{t_{s}\left({\mathbf{p}},\tilde{{\mathbf{p}}};E-\frac{3}{4m}q^{2} \right)}
{E-\frac{\tilde{p}^{2}}{m}-\frac{3}{4m}q^{2}}
\langle {\tilde{\mathbf{p}}}{\mathbf{q}}|V_{4}(1+P)|\psi\rangle
 \Biggr ] \nonumber \\
\label{eq:2.10}
\end{eqnarray}
A three-body force (3BF) with two  scalar meson exchanges and a 
constant meson-nucleon amplitude
can be written in the form of  Eq.~(\ref{eq:2.2}) with
\begin{eqnarray}
V_4 \equiv V_4^{(1)}\propto\frac{F(Q^{2})}{Q^{2}+m^{2}_{s}} \: 
\frac{F(Q'^{2})}{Q'^{2}+m^{2}_{s}}
\label{eq:2.11}
\end{eqnarray}
and a cutoff function
\begin{equation}
F(Q^{2}) = \left(\frac{\Lambda^{2}-m_s^{2}}{\Lambda^{2}+Q^{2}}\right)^{2}.
\label{eq:2.12}
\end{equation}
The momentum transfer ${\mathbf{Q}} \: ({\mathbf{Q}}')$ is given by 
\begin{eqnarray}
{\mathbf{Q}}&=&{\mathbf{k}}_{3}-{\mathbf{k}}'_{3} = 
{\mathbf{p}} -{\mathbf{p}}' -
 \frac{1}{2}({\mathbf{q}}-{\mathbf{q}}')  \nonumber \\
\mathbf{Q}'&=&{\mathbf{k}}'_{2}-{\mathbf{k}}_{2} =  
{\mathbf{p}} -{\mathbf{p}}' +
\frac{1}{2}({\mathbf{q}}-{\mathbf{q}}'),
\label{eq:2.13}
\end{eqnarray}
as indicated in Fig.~\ref{fig1}.
\par
For the evaluation of  Eq.~(\ref{eq:2.10}) matrix elements of the form
$\langle{\mathbf{pq}}|V_{4}(1+P)|\psi\rangle$ need to be calculated.
From Fig.~\ref{fig1} we see that $V_4$ can be considered as a sequence of meson
exchanges in the subsystem (12), called for convenience subsystem 3, and
subsystem (31), called 2. Inserting a complete set of states of the type 3
between $V_4$ and $(1+P)|\psi\rangle$ and another complete set of states of type
2 between the two meson exchanges (see Eq.~(\ref{eq:2.11})), leads after
a straightforward evaluation  to 
\begin{eqnarray}
\langle {\mathbf{pq}}|V_{4}(1+P)|\psi \rangle &=&\int
d^{3} q'
\frac{F((-{\mathbf{p}}-\frac{1}{2}{\mathbf{q}}-{\mathbf{q}}')^{2})}
{(-{\mathbf{p}}-
\frac{1}{2}{\mathbf{q}}-{\mathbf{q}}')^{2}+m^{2}_{s}}
 \nonumber \\
&\times& \int d^{3} p'
\frac{F((-{\mathbf{p}}+\frac{1}{2}{\mathbf{q}}-
\frac{1}{2}{\mathbf{q}}'-{\mathbf{p}}')^{2})}
{(-{\mathbf{p}}+\frac{1}{2}{\mathbf{q}}-\frac{1}{2}{\mathbf{q}}'-{\mathbf{p}}')^{2}+m^{2}_{s}}
\langle {\mathbf{p}}'{\mathbf{q}}'|(1+P)|\psi\rangle  
\label{eq:2.14}
\end{eqnarray}
The propagators in Eq.~(\ref{eq:2.14}) contain linear combinations of three
or four momentum
vectors and thus the two integrations would involve magnitudes of vectors and angles
between them. 
Realizing that both meson-exchange propagators in the 3BF term
only depend on the momentum transfer in a two-body subsystem, one can rewrite
 Eq.~(\ref{eq:2.14}) as
\begin{eqnarray}
\langle{\mathbf{p}}{\mathbf{q}}|V_{4}(1+P)|\psi\rangle &=& \int
d^{3} p'd^{3} q'\
\langle{\mathbf{p}}{\mathbf{q}}|{\mathbf{p}}'{\mathbf{q}}'\rangle_{2}
\nonumber \\
&\times&\int d^{3} p''
\frac{F(({\mathbf{p}}'-{\mathbf{p}}'')^{2})}{({\mathbf{p}}'-{\mathbf{p}}'')^{2}+m^{2}_{s}}
\nonumber \\
&\times&\int d^{3} p'''d^{3} q'''\ _{2}
\langle{\mathbf{p}}''{\mathbf{q}}'|{\mathbf{p}}'''{\mathbf{q}}'''\rangle_{3}
\nonumber \\
&\times&\int
d^{3} p''''\frac{F(({\mathbf{p}}'''-{\mathbf{p}}'''')^{2})}{({\mathbf{p}}'''-{\mathbf{p}}'''')^{2}+m^{2}_{s}}\
_{3} \langle{\mathbf{p}}''''{\mathbf{q}}'''|\Psi\rangle.
\label{eq:2.15}
\end{eqnarray}
Here the subscripts $1,2,3$ of the bra and ket vectors stand for the different types
of coordinate systems described by Eq.~(\ref{eq:2.7}). Though the integrations
of  Eq.~(\ref{eq:2.15}) look more complicated, the
meson-propagators show a very simple form. In fact, the integrations over ${\mathbf{p}}''$
and ${\mathbf{p}}''''$ have an identical functional form. 
Defining
\begin{eqnarray}
F_{3}({\mathbf{p}}''',{\mathbf{q}}''') = \int
d^{3} p''''\frac{F(({\mathbf{p}}'''-{\mathbf{p}}'''')^{2})}{({\mathbf{p}}'''-{\mathbf{p}}'''')^{2}+m^{2}_{s}}\
_{3} \langle{\mathbf{p}}''''{\mathbf{q}}'''|\Psi\rangle, 
\label{eq:2.16}
\end{eqnarray}
the integration of the meson exchange between particles 2 and 1
in Eq.~(\ref{eq:2.16}) is carried out completely
in the coordinate system
of type $3$. Once $F_{3}({\mathbf{p}}''',{\mathbf{q}}''')$ is obtained, it needs to
be expressed in terms of momenta in 
a coordinate system of type 2 in order to carry out the integration
over the remaining meson exchange.  This transformation, labeled
$F_{32}({\mathbf{p}}'',{\mathbf{q}}')$  is explicitly given as 
\begin{eqnarray}
F_{32}({\mathbf{p}}'',{\mathbf{q}}') &=& \int
d^{3} p'''d^{3} q''' \ _{2}\langle
{\mathbf{p}}''{\mathbf{q}}'|{\mathbf{p}}'''{\mathbf{q}}'''\rangle_{3}
\:
F_{3}({\mathbf{p}}''',{\mathbf{q}}''') \nonumber \\
&=&F_{3}(|-\frac{1}{2}{\mathbf{p}}''-\frac{3}{4}{\mathbf{q}}'|,
|{\mathbf{p}}''-\frac{1}{2}{\mathbf{q}}'|,
\frac{(-\frac{1}{2}{\mathbf{p}}''-\frac{3}{4}{\mathbf{q}}')\cdot({\mathbf{p}}''-\frac{1}{2}{\mathbf{q}}')}
{|-\frac{1}{2}{\mathbf{p}}''-\frac{3}{4}{\mathbf{q}}'||{\mathbf{p}}''-\frac{1}{2}{\mathbf{q}}'|}).
\nonumber \\
\label{eq:2.17}
\end{eqnarray}
Here we used that $F_3({\bf p'''},{\bf q'''})$ is a scalar function due to the
total wave function $\Psi({\bf p},{\bf q})$ being a scalar in the ground state.
The integration over the second  meson exchange between particle 1 and 3
in the coordinate system of type $2$ is now given by
\begin{eqnarray}
F_{2}({\mathbf{p}}',{\mathbf{q}}')=\int
d^{3} p''\frac{F(({\mathbf{p}}'-{\mathbf{p}}'')^{2})}{({\mathbf{p}}'-{\mathbf{p}}'')^{2}+m^{2}_{s}}
F_{32}({\mathbf{p}}'',{\mathbf{q}}'). 
\label{eq:2.18}
\end{eqnarray}
The matrix element  
$\langle {\mathbf{p}}{\mathbf{q}}|V_{4} (1+P)|\psi\rangle$ is
finally obtained by integrating 
$F_{2}({\mathbf{p}}',{\mathbf{q}}')$ 
over ${\mathbf{p}}'$ and ${\mathbf{q}}'$, i.e. carrying out the final coordinate transformation
from the system of type $2$ back to the one of type $1$, 
\begin{eqnarray}
\langle{\mathbf{p}}{\mathbf{q}}|V_{4}(1+P)|\psi\rangle &=& \int
d^{3} p' d^{3} q' \langle
{\mathbf{p}}{\mathbf{q}}|{\mathbf{p}}'{\mathbf{q}}'\rangle_{2}
F_{2}({\mathbf{p}}',{\mathbf{q}}') \nonumber \\
&=&F_{2}(|-\frac{1}{2}{\mathbf{p}}+\frac{3}{4}{\mathbf{q}}|,
|-{\mathbf{p}}-\frac{1}{2}{\mathbf{q}}|,
\frac{(-\frac{1}{2}{\mathbf{p}}+\frac{3}{4}{\mathbf{q}})\cdot(-{\mathbf{p}}-\frac{1}{2}{\mathbf{q}})}
{|-\frac{1}{2}\mathbf{p}+\frac{3}{4}\mathbf{q}||-\mathbf{p}-\frac{1}{2}\mathbf{q}|}).
\nonumber \\
\label{eq:2.19}
\end{eqnarray}
Thus, the integration of Eq.~(\ref{eq:2.15}), explicitly given in separate steps
from Eq.~(\ref{eq:2.16}) to Eq.~(\ref{eq:2.19}) contains only integrations over one
vector variable at a time. It should be pointed out that 
Eqs.~(\ref{eq:2.17}) and Eq.~(\ref{eq:2.19}) are 
only three dimensional interpolations. 
Clearly, the in Eq.~(\ref{eq:2.15}) suggested
method is the preferred one for practical calculations.

The Faddeev amplitude $\psi({\mathbf{p}},{\mathbf{q}})$ is given as function of
vector Jacobi momenta and obtained as solution of a three dimensional integral
equation, Eq.~(\ref{eq:2.10}). For the ground state $\psi({\mathbf{p}},{\mathbf{q}})$
is also a scalar and thus only depends on the magnitudes of ${\mathbf{p}}$ and
${\mathbf{q}}$ and the angle between the two vectors. 
In order to solve the Eq.~(\ref{eq:2.10})
directly without introducing partial wave projection, we have to define a
coordinate system. We choose the vector ${\mathbf{q}}$ parallel to the $z$-axis
and express the remaining vectors with respect to ${\mathbf{q}}$.
For the first term in Eq.~(\ref{eq:2.10}) the relevant vectors are ${\mathbf{p}}$
and ${\mathbf{q}}'$. Thus one has aside from the magnitudes the 
following angle relations
\begin{eqnarray}
x&=&\hat{\mathbf{p}}\cdot\hat{\mathbf{q}}=\cos\theta \nonumber \\
x'&=&\hat{\mathbf{q}}'\cdot\hat{\mathbf{q}}=\cos\theta' \nonumber \\
y&=&\hat{\mathbf{p}}\cdot\hat{\mathbf{q}}'=\cos\gamma
\label{eq:2.20}
\end{eqnarray}
where
\begin{eqnarray}
\cos\gamma=\cos\theta\cos\theta'+\sin\theta\sin\theta'\cos(\phi-\phi')
=xx'+\sqrt{1-x^{2}}\sqrt{1-x'^{2}}\cos\phi'
\label{eq:2:21}
\end{eqnarray}
Since the $\phi'$ integration is over the full $2\pi$ interval we have a 
freedom of choice for the azimuthal angle angle and set $\phi=0$. With these
choice of variables  Eq.~(\ref{eq:2.10}) with only the first term, i.e. two-body
forces alone, was solved successfully in 
Ref\cite{elster98-1}. 
For the evaluation of the second term in Eq.~(\ref{eq:2.10}),
\begin{eqnarray}
\langle {\mathbf{pq}}|V_{4}(1+P)|\psi\rangle =\langle {\mathbf{pq}}|V_{4}|\Psi\rangle,
\label{eq:2.22} 
\end{eqnarray}
we start with calculating first $F_{3}({\mathbf{p}}''', {\mathbf{q}}''')$, 
Eq.~(\ref{eq:2.16}), and realize, that for this integration we can choose 
${\mathbf{q}}'''$ parallel to the $z$-axis with corresponding simplifications for one of the azimuthal
angles.
This leads to the explicit expression 
\begin{eqnarray}
\lefteqn {F_{3}(p''',q''',x''') = }   \nonumber \\ 
& & \int^{\infty}_{0}
dp''''p''''^{2}\int^{+1}_{-1}dx''''\int^{2\pi}_{0}d\phi''''
\frac{\left( \frac{\Lambda^{2}-m^{2}_{\alpha}}{\Lambda^{2}+(p'''^{2}+p''''^{2}-2p'''p''''y'''')^{2}} \right)^{2}} 
{(p'''^{2}+p''''^{2}-2p'''p''''y'''')^{2}+m^{2}_{\alpha}}
\Psi(p'''',q''',x''''). \nonumber \\
\label{eq:2.24}
\end{eqnarray}
The evaluation of $F_{32}({\mathbf{p}}'',{\mathbf{q}}')$, Eq.~(\ref{eq:2.17}) is
not an integration but rather a three dimensional interpolation and explicitly given by
\begin{eqnarray}
\lefteqn {F_{32}(p'',q',x'') =} \nonumber \\
& & F_{3}\left(\frac{1}{2}\sqrt{\frac{9}{4}q'^{2}+p''^{2}+3p''q'x''},
                   \sqrt{\frac{1}{4}q'^{2}+p''^{2}- p''q'x''},
        \frac{\frac{3}{8}q'^{2}-\frac{1}{2}p''^{2}-\frac{1}{2}p''q'x''}
        {\left|-\frac{3}{4}{\mathbf{q}}'-\frac{1}{2}{\mathbf{p}}''\right|
         \left|+{\mathbf{p}}''-\frac{1}{2}{\mathbf{q}}'\right|}
       \right). \nonumber \\ 
\label{eq:2.25}
\end{eqnarray}
with
\begin{eqnarray}
\left|-\frac{3}{4}{\mathbf{q}}'-\frac{1}{2}{\mathbf{p}}''\right|&=&
\frac{1}{2}\sqrt{\frac{9}{4}q'^{2}+p''^{2}+3p''q'x''}
\nonumber \\
\left|+{\mathbf{p}}''-\frac{1}{2}{\mathbf{q}}'\right|&=&\sqrt{\frac{1}{4}q'^{2}+p''^{2}-
p''q'x''}. 
\label{eq:2.26}
\end{eqnarray}
For the interpolation we apply the cubic splines introduced in Ref\cite{hueber97-1}.
The integration over the second meson exchange, i.e. the calculation of 
$F_{2}({\mathbf{p}}',{\mathbf{q}}')$ of Eq.~(\ref{eq:2.18}) is functionally the same as 
Eq.~(\ref{eq:2.24}), since we can choose the variable ${\mathbf{q}}'$ parallel to the 
$z$-axis. Thus we have the same expression as Eq.~(\ref{eq:2.24}) with 
$p'$, $q'$, $x'$, $p''$, $x''$ and $\phi''$
instead of $p'''$, $q'''$, $x'''$, $p''''$, $x''''$ and $\phi''''$. Finally, the
matrix element $\langle {\mathbf{pq}}|V_{4}|\Psi\rangle$ is explicitly 
obtained by a second interpolation as
\begin{eqnarray}
\langle{\mathbf{p}}{\mathbf{q}}|V_{4}|\Psi\rangle& \equiv &
V_{4}\Psi(p,q,x) \nonumber \\
&=&F_{2}\left(\frac{1}{2}\sqrt{\frac{9}{4}q^{2}+p^{2}-3pqx},
                   \sqrt{\frac{1}{4}q^{2}+p^{2}+ pqx},
        \frac{-\frac{3}{8}q^{2}+\frac{1}{2}p^{2}-\frac{1}{2}pqx}
        {\left|+\frac{3}{4}{\mathbf{q}}-\frac{1}{2}{\mathbf{p}}\right|
         \left|-{\mathbf{p}}-\frac{1}{2}{\mathbf{q}}\right|}
       \right)\nonumber \\ \label{eq:2.27}
\end{eqnarray}
The last term of Eq.~(\ref{eq:2.10}) requires an additional integration
of the matrix element $\langle {\mathbf{p}}{\mathbf{q}}|V_{4}(1+P)|\psi\rangle$  
and the half shell two body $t$-matrix. Again,
with choosing ${\mathbf{q}}$ parallel to the $z$-axis we only have 
three vectors to consider, $\tilde{\mathbf{p}}$, ${\mathbf{p}}$ 
and ${\mathbf{q}}$, thus the integration is of a similar type as
the one of the first term in Eq.~(\ref{eq:2.10}),
\begin{eqnarray}
&&\frac{1}{2}\int d^{3}\tilde{p}
\frac{t_{s}\left({\mathbf{p}},\tilde{{\mathbf{p}}};E-\frac{3}{4m}q^{2} \right)}
{E-\frac{\tilde{p}^{2}}{m}-\frac{3}{4m}q^{2}}
\langle {\tilde{\mathbf{p}}}{\mathbf{q}}|V_{4}(1+P)|\psi\rangle \nonumber \\
&=&\frac{1}{2}
\int^{\infty}_{0} d\tilde{p}\tilde{p}^{2} \int^{+1}_{1}d\tilde{x}
\int^{2\pi}_{0}d\tilde{\phi}
\ \frac{t_{s}\left(p,\tilde{p},\tilde{y_{p}},;E-\frac{3}{4m}q^{2} \right)}
{E-\frac{\tilde{p}^{2}}{m}-\frac{3}{4m}q^{2}}
V_{4}\Psi(\tilde{p},q,\tilde{x}) 
\label{eq:2.29}
\end{eqnarray}
with
\begin{eqnarray}
\tilde{x}&=&\hat{\tilde{\mathbf{p}}}\cdot\hat{\mathbf{q}}\nonumber \\
\tilde{y}_{p}&=&\hat{\tilde{\mathbf{p}}}\cdot\hat{\mathbf{p}}=
x\tilde{x}+\sqrt{1-x^{2}}\sqrt{1-\tilde{x}^{2}}\cos\tilde{\phi}.
\label{eq:2.30}
\end{eqnarray}
We obtain the energy eigenvalue $E$ and Faddeev component $\psi(p,q,x)$ of 
the three-body system by solving 
Eq.~(\ref{eq:2.10}). 

Finally, we want to give the explicit expression for the full wave
function from Eq.~(\ref{eq:2.6}), which is 
\begin{eqnarray}
\Psi(p,q,x)&=&\psi(p,q,x) \nonumber \\
&+&
\psi\left( 
\frac{1}{2}\sqrt{ \frac{9}{4}q^{2}+p^{2}+3pqx},
           \sqrt{ \frac{1}{4}q^{2}+p^{2}-pqx },  
\frac{\frac{3}{8}q^{2}-\frac{1}{2}p^{2}-\frac{1}{2}pqx}
     {\left|-\frac{3}{4}{\mathbf{q}}-\frac{1}{2}{\mathbf{p}} \right|
        \left|-\frac{1}{2}{\mathbf{q}}+{\mathbf{p}} \right| } 
\right) \nonumber \\
&+&
\psi\left( 
\frac{1}{2}\sqrt{ \frac{9}{4}q^{2}+p^{2}-3pqx},
           \sqrt{ \frac{1}{4}q^{2}+p^{2}+pqx},  
\frac{-\frac{3}{8}q^{2}+\frac{1}{2}p^{2}-\frac{1}{2}pqx}
     {\left|+\frac{3}{4}{\mathbf{q}}-\frac{1}{2}{\mathbf{p}} \right|
        \left|-\frac{1}{2}{\mathbf{q}}-{\mathbf{p}} \right| } 
\right). \nonumber \\
\label{eq:2.31}
\end{eqnarray}
The wave function is normalized according to
\begin{eqnarray}
\int d^{3}p\:d^{3}q\:\Psi^{2}({\mathbf{p}},{\mathbf{q}})=1.
\label{eq:2.33}
\end{eqnarray}


\section{Calculation of the three-body Bound State}
Our model calculations are based on
Yukawa interactions. As two-body force (2BF) we 
employ a Malfliet-Tjon \cite{malfliet69-1} type  potential
\begin{eqnarray}
V({\mathbf{q}},{\mathbf{q}}')=
&-&\frac{g^2_A}{(2\pi)^3}\frac{1}{({\mathbf{q}}-{\mathbf{q}}')^{2}+m^{2}_{A}}
\left(\frac{\Lambda^{2}_{A}-m^{2}_{A}}{({\mathbf{q}}-{\mathbf{q}}')^{2}+
\Lambda^{2}_{A}}\right)^{2}
\nonumber \\
&+&\frac{g^2_R}{(2\pi)^3}\frac{1}{({\mathbf{q}}-{\mathbf{q}}')^2+m^2_R}
\left(\frac{\Lambda^2_R-m^2_R}{({\mathbf{q}}-{\mathbf{q}}')^2+\Lambda^2_R}\right)^{2},
\label{eq:3.1}
\end{eqnarray}
which is modified by a cutoff function of dipole type. With this choice, the 2BF
and the 3BF have similar functional forms for the scalar meson exchanges. The force
in Eq.(\ref{eq:3.1}) is a superposition of a short-ranged repulsive and 
long-ranged attractive 
Yukawa interactions. The coupling constants and exchanged meson masses are characterized 
by subscripts R and A respectively. The exchanged masses $m_{A}$ and $m_{R}$ are those 
from the original Malfliet-Tjon model, the coupling constants are chosen so that the
two-body force gives a binding of the three-body system which is slightly smaller than 
the experimental value of the triton binding, which is $8.48\:{\mathrm{MeV}}$. 
The cutoff masses
$\Lambda$ have values typical for one-boson-exchange models. 
The parameters for the 2BF,
named MT2-II, are given in Table~\ref{table1}. They lead to a two-body binding energy of 0.284 MeV, and
the S-wave phase-shift roughly follows the shape of the 
experimental one for the state $^3S_1$, though being less attractive. Since we
neglect spin degrees of freedom, any closer adjustment to real phase shifts would be meaningless
anyhow.

With this interaction we first solve the Lippmann-Schwinger equation for 
the fully-off-shell
two-body $t$-matrix directly as function of vector variables as described in 
Ref.~\cite{elster98-2}.
This $t$-matrix is then symmetrized to obtain $t_{s}(p',p,x,E-\frac{3}{4m}q^{2})$.   
The eigenvalue equation, Eq.~(\ref{eq:2.10}), for the three-body bound state is solved 
iteratively by a Lanczo's type algorithm described in detail in Ref.\cite{stadler}.
Using the 2BF alone, the binding energy of the three-body system is calculated 
as $E=7.6986\:{\mathrm{MeV}}$.
\par 
The simplest 3BF we want to apply in our study has the functional form
\begin{eqnarray}
V_{4}=\frac{1}{(2\pi)^{6}} \frac{a_{\alpha}}{m_{\alpha}} g^{2}_{\alpha}
\frac{F_{\alpha}(Q^{2})}{Q^{2}+m^{2}_{\alpha}}
\frac{F_{\alpha}(Q'^{2})}{Q'^{2}+m^{2}_{\alpha}}, 
\label{eq:3.2}
\end{eqnarray}
where
\begin{eqnarray}
F_{\alpha}(Q^{2})&=&
\left(\frac{\Lambda^{2}_{\alpha}-m^{2}_{\alpha}}{\Lambda^{2}_{\alpha}+Q^{2}}\right)^{2}.
\label{eq:3.3}
\end{eqnarray}
Choosing $a_{\alpha}$ to be a negative constant makes this force purely attractive. 
The parameters of this force, which we name MT3-I in the following, 
are given in Table~\ref{table2}. They are chosen to give a small attractive contribution 
to the three-body binding energy, such that we end up in the neighborhood of the triton binding
energy. For the force parameters in Table~\ref{table2} 
we obtain $E=8.8732$~MeV. 

In order to solve the eigenvalue equation, Eq.~(\ref{eq:2.10}), for the Faddeev component
$\psi(p,q,x)$, we use Gaussian grid points in $p$, $q$, and $x$. 
The momentum and angle grids
in the integrations of the 3BF in Eqs.~(\ref{eq:2.24}) and (\ref{eq:2.29})
have the same sizes as the ones for $p$, $q$, and $x$.
This is very reasonable, since the integrations over the meson exchange contributions of
the 3BF, Eq.~(\ref{eq:2.24}), require  grids similar in range to the one used to
calculate the two-body t-matrix. The p-grid is defined between 0 and $p_{max}=60 fm^{-1}$,
whereas for the q-grid a maximum value $q_{max} = 40 fm^{-1}$ is sufficient. For the
angle ($x$) integration, the preferred number is 42 grid points. Further
details concerning the grid choices are given in Ref.~\cite{elster98-1}. 
Typical grid sizes are $97 \times 97 \times 42$ to obtain an accuracy  
in the binding energy of 5 significant figures. The convergence of the three-body binding
energy $E$ as function of the number of grid points is shown in Table~III, 
where we see a convergence of the energy eigenvalue and the expectation values within
5 digits. 

The three-body wave function is calculated from the Faddeev component
using Eq.~(\ref{eq:2.31}). Since this wave function enters the eigenvalue equation,
Eq.~(\ref{eq:2.10}), we also need to worry about the quality of the calculation
of $\Psi(p,q,x)$. One check of the overall quality of the wave function
 is a comparison of the expectation
value of the total Hamiltonian $\langle H \rangle$ with the calculated value $E$ from
the solution of the Faddeev equation. Explicitly, we evaluate
\begin{eqnarray}
\langle H \rangle
 \equiv \langle\Psi|H|\Psi\rangle=\langle\Psi|H_{0}|\Psi\rangle+\langle\Psi|V_{II}|\Psi\rangle+
\langle\Psi|V_{123}|\Psi\rangle,
\label{eq:3.4}
\end{eqnarray}
where $V_{II}$ represents the 2BF $\sum^{3}_{i=1}V^{i}$ and $V_{123}$
the three-body defined in Eq.~(\ref{eq:2.2}).    
The expectation value of the kinetic energy $\langle H_0 \rangle$ and the two-body potential
energy $\langle V_{II} \rangle$ are given as \cite{elster98-1}
\begin{eqnarray}
\langle H_0 \rangle& \equiv &
\langle\Psi|H_{0}|\Psi\rangle=3\langle\psi|H_{0}|\Psi\rangle
\nonumber \\
&=& 3\cdot
8\pi^{2}\int^{\infty}_{0}p^{2}dp\int^{\infty}_{0}q^{2}dq\left(
\frac{p^{2}}{m}+\frac{3q^{2}}{4m} \right)\int^{+1}_{-1}dx
\psi(p,q,x)\Psi(p,q,x) \nonumber \\
\label{eq:3.5}
\end{eqnarray}
and
\begin{eqnarray}
\langle V_{II} \rangle & \equiv &
\langle\Psi|V_{II}|\Psi\rangle=3\langle\Psi|V^{1}|\Psi\rangle
\nonumber \\
&=&3\cdot
8\pi^{2}\int^{\infty}_{0}p^{2}dp\int^{\infty}_{0}q^{2}dq\int^{+1}_{-1}dx\int^{\infty}_{0}p'^{2}dp'
\int^{+1}_{-1}dx'
\nonumber \\
&\times&\Psi(p,q,x)v_{1}(p,p',x,x')\Psi(p',q,x')
\label{eq:3.6}
\end{eqnarray}
where
\begin{eqnarray}
v_{1}(p,p',x,x')=\int^{2\pi}_{0}d\phi
V^{1}(p,p',xx'+\sqrt{1-x^{2}}\sqrt{1-x'^{2}}\cos\phi).
\label{eq:3.7}
\end{eqnarray}
The expectation value of the three-body potential energy, $\langle V_{123} \rangle$, 
is given by 
\begin{eqnarray}
\langle V_{123} \rangle& \equiv &
\langle\Psi|V_{123}|\Psi\rangle=3\langle\Psi|V_{4}|\Psi\rangle
\nonumber \\
&=&3\cdot
8\pi^{2}\int^{\infty}_{0}p^{2}dp\int^{\infty}_{0}q^{2}dq\int^{+1}_{-1}dx
\Psi(p,q,x)V_{4}\Psi(p,q,x).
\label{eq:3.8}
\end{eqnarray}
Here the integrations need the evaluation of the matrix element 
$\langle{\mathbf{p}}{\mathbf{q}}|V_{4}|\Psi\rangle$ of Eq.~(\ref{eq:2.14}). The expectation
values of the kinetic and potential energies are listed in Table~III as functions of
the size of the $p-q-x$ grid.  Table~III
shows that the expectation values given above 
converge for 5 significant figures when the grid sizes for $p$ and $q$ exceed $77$ points.
However, Table~III
also shows that the expectation value $\langle H \rangle$ does not exactly 
converge to the calculated value $E$ despite increased grid size. The difference between
the two quantities is of the order of 10 keV. This behavior is in 
contrast to calculations based solely on two-body forces. 
Calculations with $V_{123}=0$
are shown in Table~\ref{table4} as function of the size of the  $p-q-x$ grid. 
Here the convergence
of $\langle H \rangle$ to the calculated value $E$ as function of the grid size is 
much better than in the case where an attractive 3BF is included. 

It is well known that three nucleon forces based on multi-meson exchanges can have
attractive as well as repulsive pieces. Thus we also consider a model of this type
given as
\begin{eqnarray}
V_{4}&=&\frac{1}{(2\pi)^{6}}\frac{a_{\alpha}}{m_{\alpha}} g^{2}_{\alpha}
\frac{F_{\alpha}(Q^{2})}{Q^{2}+m^{2}_{\alpha}}
\frac{F_{\alpha}(Q'^{2})}{Q'^{2}+m^{2}_{\alpha}}\nonumber \\
&+&\frac{1}{(2\pi)^{6}}\frac{a_{\alpha\rho}}{\sqrt{m_{\alpha}m_{\rho}}}
g_{\alpha}g_{\rho} \left(
\frac{F_{\alpha}(Q^{2})}{Q^{2}+m^{2}_{\alpha}}
 \frac{F_{\rho}(Q'^{2})}{Q'^{2}+m^{2}_{\rho}}+
 \frac{F_{\rho}(Q'^{2})}{Q'^{2}+m^{2}_{\rho}}
 \frac{F_{\alpha}(Q^{2})}{Q^{2}+m^{2}_{\alpha}} \right).
\label{eq:4.2}
\end{eqnarray}
Here the first term represents an attractive force, characterized by a negative
coupling
$a_{\alpha}$, whereas the second term represent a repulsive force, i.e.
$a_{\alpha\rho}$
is positive. Since the masses of the exchange mesons are different, the form of the
second, repulsive term guarantees that $V_4$ is symmetric under a permutation of
nucleons
2 and 3. The cutoff functions $F_{\alpha}$ and $F_{\rho}$ have the same functional
form as
given in Eq.~(\ref{eq:3.3}). The parameters of this 3BF, named MT3-II in
the
following, are given in Table~\ref{table5}. They are chosen so that the correction due
to this
3BF to the three-body binding energy calculated with the 2BF
MT2-II is small. The binding energy $E$ with this MT3-II 3BF gives $E =
8.6478$~MeV. 

The expectation values of the kinetic and potential energies are listed in
Table~\ref{table6a} as functions of the size of the $p-q-x$ grid. Again, the expectation
values converge within five significant figures when the grid sizes for $p$ and $q$
exceed 77 points. However, now the difference of the expectation value for the total
Hamiltonian $\langle H \rangle$ deviates from the calculated eigenvalue $E$ only by
5~keV, a number being similar to calculations carried out in a partial wave decomposition
and based on realistic forces \cite{NoggaPhd}.

All these numbers are not meant to provide insight into the physics of three interacting nucleons, but
serve only as a demonstration that this technique allows a very accurate and easy handling of typical
nuclear forces consisting of attractive and repulsive (short range) parts. In addition, they will
serve as benchmarks for future studies.


\section{Wave Function Properties}

Despite the fact that there is no spin dependence and questions about effects of realistic forces can
not be posed, we want to display some wave function properties, which are often studied in the context
of realistic forces. It will turn out that qualitatively they also appear in our simple three boson
model. The probability of finding a nucleon with momentum $q$ in the nucleus is given as
\begin{equation}
n(q)= 2\pi q^2 \int^{\infty}_{0} dp \: p^2 \int^{+1}_{-1}dx \Psi^{2}(p,q,x) .
\label{eq:4.1}
\end{equation}
The total wave function $\Psi(p,q,x)$ is given by Eq.~({\ref{eq:2.31}). 
In Fig. 2 we show $n(q)$ with and without three-body forces. There is hardly any change, a fact with
has been noticed before in the context of realistic nuclear forces \cite{NoggaPhd}. 
In addition, the shoulder of
the distribution around 2-4~fm$^{-1}$ is qualitatively similar to the case when using realistic
forces. The momentum distribution $n(q)$ is shown in Fig.~\ref{fig2} for three different cases,
one based on a calculation with two-body forces alone (dotted line), and the other two for which the
two different 3BFs are included. First we notice that the two 3BF's, though different in character, lead
to essentially the same momentum distribution. Compared to the momentum distribution given by
the two-body force alone, the minimum is shifted to a slightly higher momentum.

Another property often investigated is the probability to find two nucleons at a distance $r$. 
To obtain that quantity we generate the total wave function in configuration space as 
\begin{equation}
\Psi({\mathbf{r,R}})=\int d^{3}p\ d^{3}q\
\Psi({\mathbf{p,q}}) \exp(i{\mathbf{p}}\cdot{\mathbf{r}})
\exp(i{\mathbf{q}} \cdot {\mathbf{R}}).
\label{eq:4.3}
\end{equation}
Here $\Psi({\bf p}$, ${\bf q})$ is the total wave function in
momentum space as  given by Eq.~(\ref{eq:2.6}). The variables ${\bf r}$ and
${\bf R}$ are conjugate to the Jacobi momenta ${\bf p}$, ${\bf q}$ and
given as
\begin{eqnarray}
{\mathbf{r}}&=&{\mathbf{x}}_{2}-{\mathbf{x}}_{3}, \nonumber \\
{\mathbf{R}}&=&{\mathbf{x}}_{1}-\frac{1}{2}({\mathbf{x}}_{2}+{\mathbf{x}}_{3}).
\label{eq:4.4}
\end{eqnarray}
where ${\bf x}_1$, ${\bf x}_2$ and ${\bf x}_3$ are the
coordinates of three nucleons in configuration space \cite{gloeckle-book}.

For the explicit calculation of the double Fourier transformation we first consider
the ${\bf q}$-integration
\begin{eqnarray}
\int d^{3}q\exp(i\mathbf{q}\cdot\mathbf{R})\Psi(\mathbf{p,q}).
\end{eqnarray}
We choose the vector ${\bf p}$ parallel to the z-axis and define the angles
$\hat{\mathbf{q}}\cdot\hat{\mathbf{z}}=x_{q}$ and $\hat{\mathbf{R}}\cdot\hat{\mathbf{z}}=x_{R}$.
Since the integration is carried out over all space, we can set $\phi_{R}=0$, and
obtain
\begin{eqnarray}
&&\int d^{3}q\exp(i\mathbf{q}\cdot\mathbf{R})\Psi(\mathbf{p,q})
\nonumber \\
&&=\int^{\infty}_{0}q^{2}dq\int^{+1}_{-1}dx_{q}\int^{2\pi}_{0}d\phi_{q}\exp(iqR\hat{\mathbf{R}}\cdot\hat{\mathbf{q}})\Psi(p,q,x_{q}),
\end{eqnarray}
where
\begin{eqnarray}
\hat{\mathbf{R}}\cdot\hat{\mathbf{q}}=x_{q}x_{R}+\sqrt{1-x^{2}_{q}}\sqrt{1-x^{2}_{R}}\cos\phi_{q}.
\end{eqnarray}
Thus, the integration over $\phi_{q}$ can be carried out separately
\begin{eqnarray}
&&\int^{2\pi}_{0}d\phi_{q}\
\exp(iqR\sqrt{1-x^{2}_{q}}\sqrt{1-x^{2}_{R}}\cos\phi_{q})\nonumber
\\
&&=2\pi J_{0}(qR\sqrt{1-x^{2}_{q}}\sqrt{1-x^{2}_{R}}\cos\phi_{q}).
\end{eqnarray}
Summarizing the above leads to the intermediate result
\begin{eqnarray}
&&\int d^{3}q\exp(i\mathbf{q}\cdot\mathbf{R})\Psi(\mathbf{p,q})
\nonumber \\
&&=2\pi\int^{\infty}_{0}q^{2}dq\int^{+1}_{-1}dx_{q}
J_{0}(qR\sqrt{1-x^{2}_{q}}\sqrt{1-x^{2}_{R}}\cos\phi_{q})\exp(iqRx_{q}x_{R})\Psi(p,q,x_{q})
\nonumber \\
&&\equiv 2\pi \: \Psi_{p}(p,R,x_{R}).
\end{eqnarray}
Next, we consider the integration over ${\bf p}$, where it is convenient to choose 
the vector ${\bf r}$
parallel to the z-axis. Thus, the following angle, $\hat{\mathbf{R}}\cdot\hat{\mathbf{z}} 
\equiv \hat{\mathbf{R}}\cdot\hat{\mathbf{r}} \equiv x_{R}$,
needs to be considered, and the integration over
$x_{p}$ and $\phi_{p}$ can be carried out separately as
\begin{eqnarray}
\int^{+1}_{-1}dx_{p}\int^{2\pi}_{0}d\phi_{p}\exp(iprx_{p})=4\pi\frac{\sin(pr)}{pr}
\end{eqnarray}
Finally, the Fourier transform of $\Psi(p,q,\hat{\mathbf{p}}\cdot\hat{\mathbf{q}})$
can be calculated as
\begin{equation}
\Psi(r,R,x_R)=
\frac{8\pi^{2}}{r}\int^{\infty}_{0}dp\left[\sin(pr)p\Psi_{p}(p,R,x_{R})\right],
\label{eq:4.5}
\end{equation}
where
\begin{equation}
\Psi_{p}(p,R,x_{R})=\int^{\infty}_{0}q^{2}dq\int^{+1}_{-1}dx_{q}
J_{0}\left(qR\sqrt{1-x^{2}_{q}}\sqrt{1-x^{2}_{R}}\right)\cos(qRx_{q}x_{R})\Psi(p,q,x_{q}).
\label{eq:4.6}
\end{equation}
With this the two-body correlation function $c(r)$ is defined as
\begin{equation}
c(r)=2\pi
r^{2}\int^{\infty}_{0}dR \: R^2 \int^{+1}_{-1}dx_{R}\Psi^{2}(r,R,x_{R}).
\label{eq:4.7}
\end{equation}
The correlation function $c(r)$ describes the probability to find two nucleons within a
relative distance $r$. In Fig. 3 the correlation functions are displayed based on a 
calculation with two-body forces (dotted line) and based on calculations with the two different 3BF's. 
Though our model is very simple, the functions $c(r)$ are similar to the ones obtained with 
realistic forces \cite{NoggaPhd}. We see that the maximum of $c(r)$ is shifted slightly to a smaller
value of $r$ once a 3BF is included, which is consistent with the minimum of $n(q)$ being shifted to a
slightly higher momentum. The position of the maximum of $c(r)$ does not depend on the type of 3BF,
however the actual shape of the function does. 

Since our three-body system consists of three identical nucleons acted on by scalar
forces, the three nucleons form a ground state where the most probable positions of the nucleons
have the 
shape of an equilateral triangle. The expectation values of the Jacobi coordinates $r$ and
$R$ can be calculated as
\begin{eqnarray}
\langle r \rangle =
\int^{\infty}_{0}R^{2}dR\int^{\infty}_{0}r^{2}dr\int^{+1}_{-1}dx_{R}
\ r\Psi^{2}(r,R,x_{R}), \nonumber \\
\langle R \rangle =
\int^{\infty}_{0}R^{2}dR\int^{\infty}_{0}r^{2}dr\int^{+1}_{-1}dx_{R} \
R\Psi^{2}(r,R,x_{R}).
\label{eq:4.8}
\end{eqnarray}
Here the values  $\langle r \rangle$ and $\langle R \rangle$ are the 
length and height of the equilateral triangle. The geometrical relation between the length
and height of an equilateral triangle is given by
\begin{equation}
\langle r \rangle:\langle R \rangle=2 : \sqrt{3}.
\label{eq:4.9}
\end{equation}
We also define a deviation $\delta$ by
\begin{equation}
\delta=\frac{\langle r \rangle/\langle R \rangle-2/\sqrt{3}}{2/\sqrt{3}}\times 100.
\label{eq:4.10}
\end{equation}
to account for deviations from an ideal geometric triangle. 
In Table~\ref{tabledel} we show the corresponding values for $\langle r \rangle$, $\langle R
\rangle$, and $\delta$ for the MT2-II 2BF alone and the cases where the two different 3BF
discussed above are added. In all cases listed, the deviation from an ideal equilateral
triangle is 3\% or less, also indicating that our calculations are very accurate.

We would like to add a little excursion into a playground with forces. Assume there are only
three-body forces. Can one generate a three-body wave function which has about the same binding
energy, single nucleon momentum distribution, and two-body correlation function as given by two-body
forces alone? This is indeed possible. For that aim we have chosen purely attractive two- and
three-body forces, the parameters of which are given in Table~\ref{table11}. The binding energies and
expectation values of kinetic and potential energies for the two cases are displayed in
Table~\ref{table7}. It
turns out that also $n(q)$ and $c(r)$ are close to each other as shown in Figs. 3 and 4. 
We do not know of a physical realization of such a scenario with pure three-body forces, but maybe the
reader may find that little excursion equally entertaining as we do.


\section{Summary and Outlook}

We derived and calculated  the Faddeev equation for three identical bosons
interacting by two- and three-body forces. The equation is
formulated in momentum space directly in terms of momentum vectors, i.e.
without angular momentum decomposition. It is demonstrated that
this equation can be solved by integrating over magnitudes
of momenta and various angles. In doing so we encounter interpolations
which are carried out by cubic Hermitian splines \cite{hueber97-1}. The Faddeev equation
is solved by iteration using a Lanczo's type method \cite{stadler,saake}. An accuracy
of 5 digits in the energy eigenvalue can easily be achieved. In
comparison to an angular momentum decomposition which is commonly used
\cite{gloeckle-book},
this direct approach has great advantages. It avoids the very
 involved angular momentum algebra  occurring for
 the permutations and especially  for
the  three-body forces \cite{coon,hueber97-1}.

 All two- and three-body forces we employ are of meson exchange type, either purely
 attractive or attractive and repulsive. The mesons responsible
 for the attraction have masses around 300 MeV , the ones for the repulsion
 around 600 MeV. These forces, serving as a reference,
  are chosen with a view towards nuclear physics. Thus the reference two-body
 forces lead to a three-body binding energy somewhat smaller than
 8.48 MeV (the experimental $^3$H binding energy), and the reference
 three-body forces add about 1 MeV additional binding energy.

   For these type of forces we evaluated the single nucleon momentum
 distributions and the two-body correlation functions.
 These quantities turned out to be qualitatively very similar to what is
 achieved with realistic spin dependent forces \cite{NoggaPhd,NoggaPL}. 

   We also evaluated the expectation values $\langle r \rangle$ for a pair distance and 
$\langle R \rangle$ for
 the distance of a third particle to the c.m. of the corresponding
 pair. In case of an equilateral triangle the ratio of these two
 quantities is $2/\sqrt{3}$.  Since we have three identical bosons
 the position of the three particles in the ground state 
should be with highest probability 
at the corners of an equilateral
 triangle. Indeed the corresponding ratio for the expectation values
$\langle r \rangle$ and $\langle R \rangle$ turned out to be $2/\sqrt{3}$ with an
 numerical error of about 1-2\%.

  Including spin (and isospin) degrees of freedom is an additional
 task for the future, which will increase the space of states
  and will lead to
 coupled equations, but only a strictly finite number of equations.
 The form in which they will appear will depend on the way the
  spin degrees of
 freedom will be incorporated. One possibility will be the extension of  the helicity
 formalism chosen for the NN system in  a three-dimensional
  notation \cite{imam1}
 to three nucleons. The other possible  extension of the here presented formulation
is the incorporation of relativity in the instant form of
 dynamics \cite{coester}. The momentum space formulation seems ideal for
 that. First steps have been already undertaken \cite{kamada,kamada2}. Since
 relativity will be of importance at higher energies our treatment
 with momentum vectors and avoiding angular momentum decomposition will
 be awarding in that respect.


\vfill

\acknowledgements This work was performed in part under the
auspices of the U.~S.  Department of Energy under contract
No. DE-FG02-93ER40756 with Ohio University. The computational support of the 
Ohio Supercomputer Center (OSC) for the use of their facilities under Grant No.~PHS206,
the Neumann Institute for Computing (NIC) under project JIKP01 and the National Energy
Research Supercomputer Center (NERSC) is acknowledged.  The authors want to thank
A.~Nogga for helpful and stimulating discussions.



 
\begin{table}\caption{\label{table1}
The parameters of the MT2-II two-body force.} 
\begin{tabular}{cccccc}
\mbox{  }
 $g^{2}_{A}/4\pi$ & $m_{A}$[MeV] & $\Lambda_{A}$[MeV]&
 $g^{2}_{R}/4\pi$ & $m_{R}$[MeV] & $\Lambda_{R}$[MeV] \\ \hline
  3.5775 & 330.2104 & 1500.0 & 9.4086 & 612.4801 & 1500.0  \\ 
  \end{tabular}
\end{table}


\begin{table}\caption{\label{table2}The parameters of the MT3-I attractive 3BF.} 
\begin{tabular}{ccccc}
\mbox{  }
 $g^{2}_{\alpha}/4\pi$ & $m_{\alpha}$[MeV] & $\Lambda_{\alpha}$[MeV] &
  $a_{\alpha}$\\ \hline
5.0 & 305.8593 & 1000.0 & -1.73  \\ 
  \end{tabular}
\end{table}


\begin{table}\caption{\label{table3}The calculated eigenvalue $E$ from the
 the solution of the Faddeev
equation and the expectation values of the kinetic energy $\langle H_0
\rangle$, the two-body potential $\langle V_{II} \rangle$,  the
three-body potential energy $\langle V_{123} \rangle$ and the total Hamiltonian $\langle H
\rangle$ as functions of the number of grid points NP, NQ and NX for the $p-q-x$ grid. The
calculations are based on the MT2-II 2BF and the MT3-I 3BF.}
 
\begin{tabular}{cccccccc}
\mbox{  }

   NP & NQ & NX &$\langle H_0 \rangle$ (MeV)&$\langle V_{II} \rangle$ (MeV)
                &$\langle V_{123} \rangle$ (MeV)& $\langle H \rangle$ (MeV)
                & $E$ (MeV)   \\ \hline
   45 & 45 & 42 & 31.8838 & -39.4069 & -1.3392 & -8.8623  & -8.8715 \\ \hline
   61 & 45 & 42 & 31.8846 & -39.4073 & -1.3396 & -8.8623  & -8.8711 \\ \hline
   77 & 45 & 42 & 31.8848 & -39.4074 & -1.3397 & -8.8623  & -8.8709 \\ \hline
   77 & 61 & 42 & 31.8900 & -39.4133 & -1.3406 & -8.8639  & -8.8726 \\ \hline
   77 & 77 & 42 & 31.8915 & -39.4149 & -1.3409 & -8.8643  & -8.8731 \\ \hline
   87 & 87 & 42 & 31.8919 & -39.4152 & -1.3410 & -8.8644  & -8.8732 \\ \hline
   97 & 97 & 42 & 31.8920 & -39.4154 & -1.3410 & -8.8644  & -8.8732 \\ 
  \end{tabular}
  \end{table}


\begin{table}\caption{\label{table4}The calculated eigenvalue $E$ of the Faddeev
equation and the expectation values of the kinetic energy $\langle H_0
\rangle$, the two-body potential $\langle V_{II} \rangle$, and the total Hamiltonian
 $\langle H \rangle$ as functions of the number of grid points NP, NQ and NX for the
$p-q-x$ grid. The
calculations are based on the MT2-II 2BF alone.} 

\begin{tabular}{ccccccc}
\mbox{  }

   NP & NQ & NX &$\langle H_0 \rangle$ (MeV)&$\langle V_{II} \rangle$ (MeV)& 
   $\langle H \rangle$ (MeV)& $E$ (MeV)  \\ \hline
   77 & 77 & 42 & 28.6408 & -36.3390 & -7.6983 & -7.6984 \\ \hline
   87 & 87 & 42 & 28.6408 & -36.3391 & -7.6983 & -7.6984 \\ \hline
   97 & 97 & 42 & 28.6408 & -36.3392 & -7.6983 & -7.6984 \\ 
  \end{tabular}
  \end{table}


\begin{table}\caption{\label{table5} The parameters of the MT3-II 3BF.} 
\begin{tabular}{cccc}
\mbox {}
 $g^{2}_{\alpha}/4\pi$ & $m_{\alpha}$[MeV] & $\Lambda_{\alpha}$[MeV] &
  $a_{\alpha}$  \\ \hline
  5.0 & 305.8593 & 1000.0 & -2.69  \\ \hline
  $g^{2}_{\rho}/4\pi$ & $m_{\rho}$[MeV] & $\Lambda_{\rho}$[MeV] &
  $a_{\alpha\rho}$ \\ \hline
  9.0 & 650.0000 & 1900.0 &  2.40 \\ 
  \end{tabular}
\end{table}


\begin{table}\caption{\label{table6a} The calculated eigenvalue $E$ from the
 the solution of the Faddeev
equation and the expectation values of the kinetic energy $\langle H_0
\rangle$, the two-body potential $\langle V_{II} \rangle$,  the
three-body potential energy $\langle V_{123} \rangle$ and the total 
Hamiltonian $\langle H
\rangle$ as functions of the number of grid points NP, NQ and NX for the $p-q-x$ grid.
The calculations are based on the MT2-II 2BF and the MT3-II 3BF.}

\begin{tabular}{cccccccc}

\mbox{  }

   NP & NQ & NX &$\langle H_0 \rangle$ (MeV)&$\langle V_{II} \rangle$ (MeV)
                &$\langle V_{123} \rangle$ (MeV)& $\langle H \rangle$ (MeV)
                & $E$ (MeV)   \\ \hline
   45 & 45 & 42 & 31.1745 & -38.7859 & -1.0404 & -8.6518  & -8.6454 \\ \hline
   61 & 45 & 42 & 31.1767 & -38.7878 & -1.0409 & -8.6520  & -8.6456 \\ \hline
   77 & 45 & 42 & 31.1773 & -38.7881 & -1.0413 & -8.6521  & -8.6466 \\ \hline
   77 & 61 & 42 & 31.1825 & -38.7880 & -1.0478 & -8.6533  & -8.6477 \\ \hline
   77 & 77 & 42 & 31.1837 & -38.7886 & -1.0481 & -8.6530  & -8.6480 \\ \hline
   87 & 87 & 42 & 31.1842 & -38.7887 & -1.0481 & -8.6526  & -8.6478 \\ \hline
   97 & 97 & 42 & 31.1847 & -38.7892 & -1.0481 & -8.6526  & -8.6478 \\
  \end{tabular}
  \end{table}

\begin{table}\caption{\label{tabledel} The binding energy $E$, the expectation values $\langle r
\rangle$ and $\langle R \rangle$ calculated with the MT2-II 2BF alone and with the addition of
the two different 3BF's described in the text. The deviation $\delta$ characterizes the
deviation from the shape of an equilateral triangle and is
defined in Eq.~(\ref{eq:4.10}).  }
\begin{tabular}{cccc}
\mbox{}
          & MT2-II  & MT2-II + MT3-I & MT2-II + MT3-II  \\ \hline 
$E$ (MeV) &  -7.6980  & -8.873 & -8.6478 \\
$\langle r \rangle$ (fm)& 2.521  & 2.382 & 2.9401 \\
$\langle R \rangle$ (fm) & 2.221 & 2.096 & 2.5945 \\
$\delta$ (\%)   &      1.7   & 1.6  &   1.9 \\ 
  \end{tabular}
\end{table}

\begin{table}\caption{\label{table11}The parameters of the
purely attractive potentials, the 2BF MT2-I and the 3BF MMT3-I leading to the
same three-body binding energy. }
\begin{tabular}{lcccc}
\mbox{}
MT2-I &   $g^{2}_A/4\pi$ & $m_A$[MeV] & $\Lambda_A $[MeV] &  \\ \hline
   & -0.7210  & 330.2104  & 1500 & \\ \hline \hline 
MMT3-I &   $g^{2}_{\alpha}/4\pi$ & $m_{\alpha}$[MeV] &
$\Lambda_{\alpha}$[MeV] &
  $a_{\alpha}$   \\ \hline
 &  5.0000 & 60.0 & 660  & -1.90015  \\
  \end{tabular}
\end{table}

\begin{table}\caption{\label{table7}The calculated eigenvalue $E$ from the
 the solution of the Faddeev
equation,  the expectation values of the kinetic energy $\langle H_0
\rangle$, and  the
potential energy $\langle V \rangle$ for the two-body force MT2-I and the three-body
force MMT3-I. 
Both forces give similar binding energies. The expectation values of $\langle R \rangle$ and
 $\langle r \rangle$ are also given for both cases.}

\begin{tabular}{cccccc}
\mbox{}
Model &    $\langle H_0 \rangle$ (MeV)&$\langle V \rangle$ (MeV) &
      $\langle R \rangle$ (fm) & $\langle r \rangle$ (fm) &E (MeV)   \\ \hline
MT2-I &  66.967  &  -74.547 & 1.592  & 1.783   & -7.5803  \\  \hline
MMT3-I  &  67.306   &  -74.895 & 1.521 & 1.698    & -7.5504   \\
  \end{tabular}
  \end{table}

\pagebreak

\noindent
\begin{figure}
\caption{Diagrammatic representation of the three-body force $V_4^{(1)}$. Here particle
(1) is singled out by the meson-nucleon amplitude described by the blob.
 The three-body force is then given according to Eq.~(\ref{eq:2.2}).
 \label{fig1}}
 \end{figure}

 \begin{figure}
 \caption{The momentum distributions $n(q)$ calculated with the MT2-II
 two-body potential (dotted line). The solid line represents the calculation of $n(q)$
 with the  MT2-II two-body potential and the MT3-I three-body potential, the dashed line
 the corresponding calculation with the MT3-II three body potential.  
 \label{fig2}}
 \end{figure}

 \begin{figure}
 \caption{The two-body correlation function $c(r)$ calculated with 
 the MT2-II two-body potential (dotted line).
 The solid line represents the calculation of $c(r)$ with the  MT2-II two-body potential and
 the MT3-I three-body potential, the dashed line
 the corresponding calculation with the MT3-II three body potential.
 \label{fig5}}
 \end{figure}

\begin{figure}
\caption{The momentum distributions $n(q)$ calculated with the attractive  
2BF MT2-I (solid line). 
The dashed curve represents the calculation with  the purely attractive
3BF, where the parameters are chosen such that the binding energy and the 
momentum distribution are similar to the one given by
the MT2-I potential. 
\label{fig14}}
\end{figure}

\begin{figure}
\caption{The two-body correlation function $c(r)$ calculated with
the  attractive 2BF MT2-I (solid line). The dashed curve represents
the calculation with  the purely attractive
3BF as used in Fig.~\ref{fig14}
  \label{fig15}}
\end{figure}

\newpage

\input{psfig}

{\bf Fig.1}
\centerline{\hspace{15mm}\psfig{file=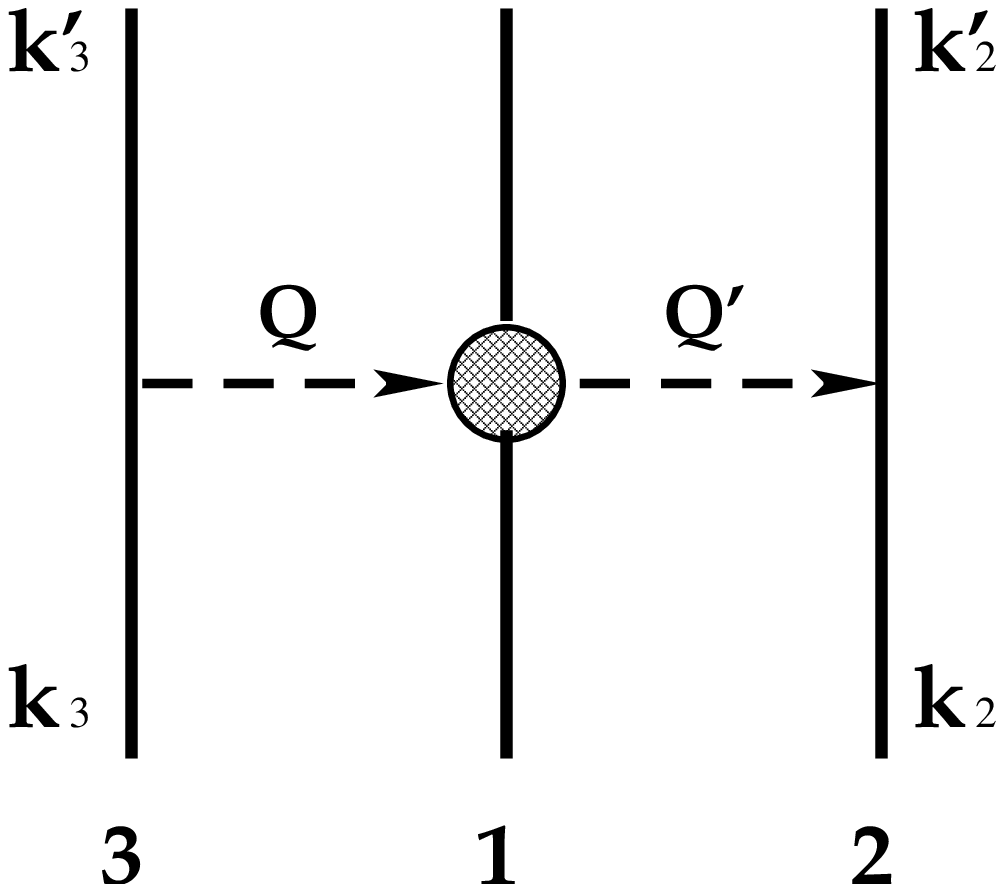,width=80mm}}

\vspace{30mm}

{\bf Fig.2}
\centerline{\hspace{15mm}\psfig{file=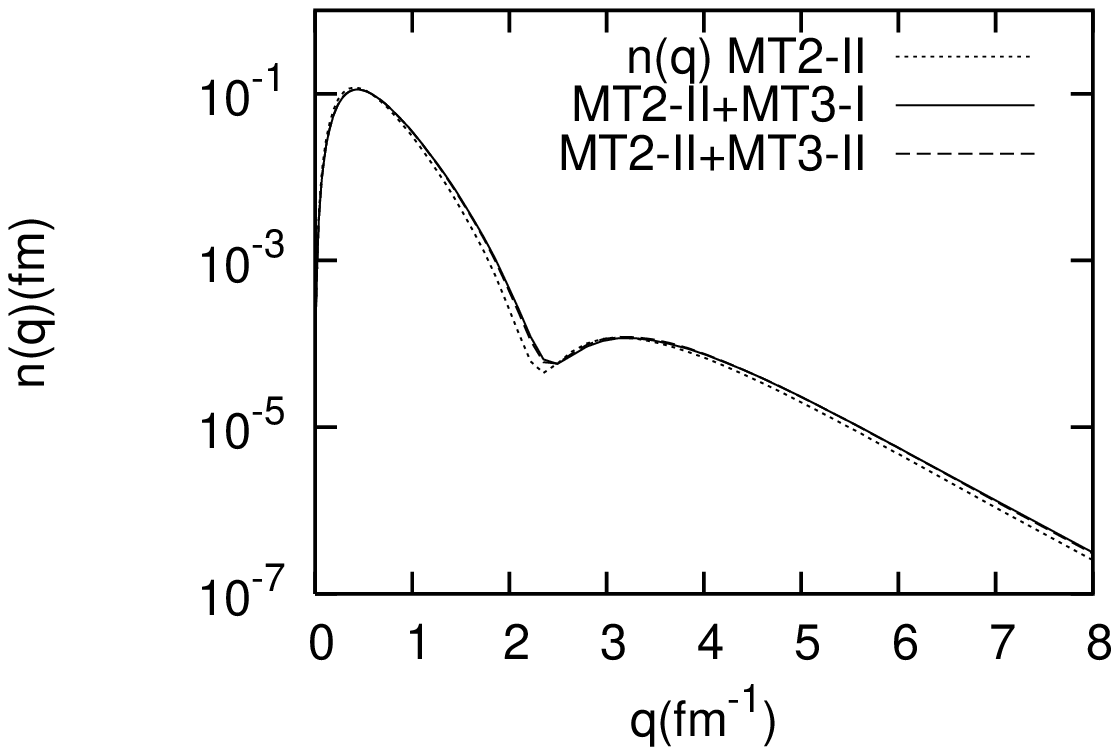,width=120mm,angle=-90}}

\newpage

{\bf Fig.3}
\centerline{\hspace{15mm}\psfig{file=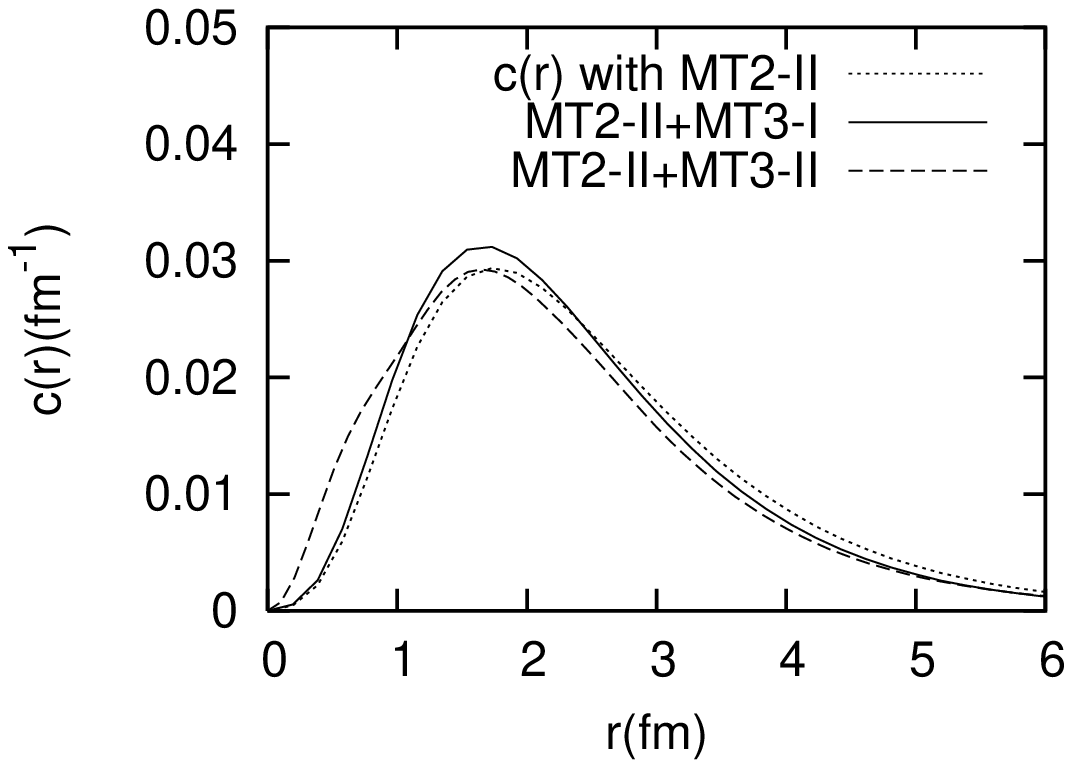,width=120mm,angle=-90}}

\vspace{10mm}

{\bf Fig.4}
\centerline{\hspace{15mm}\psfig{file=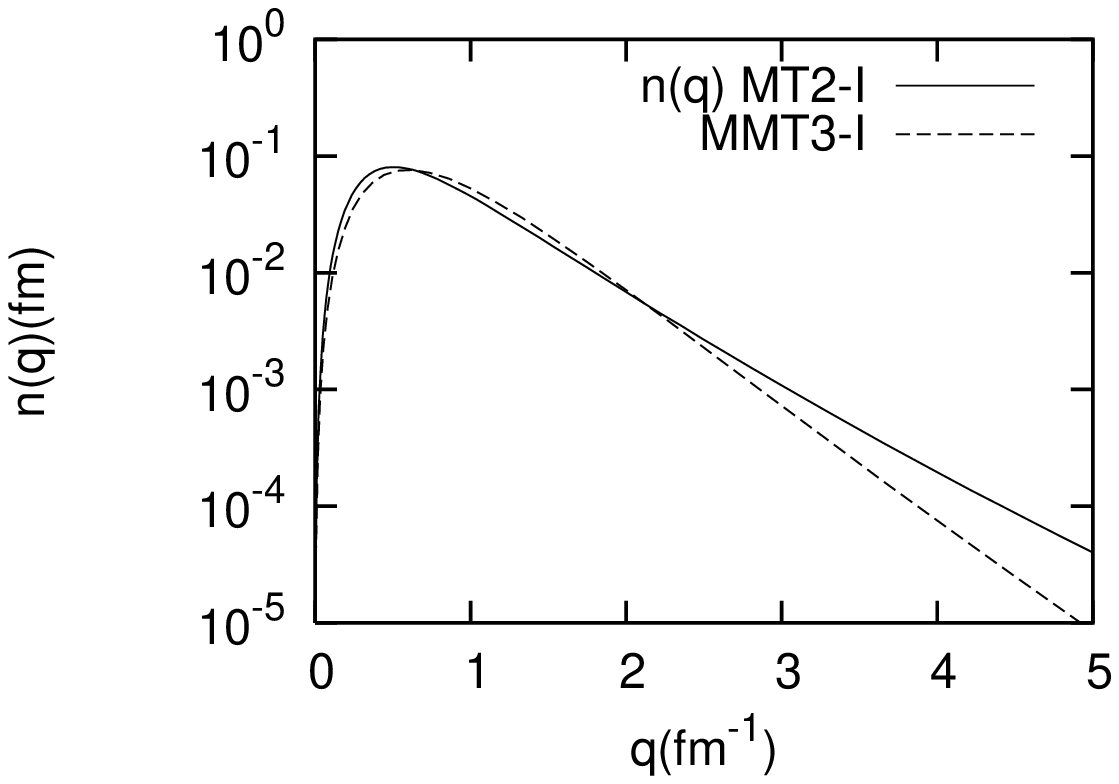,width=120mm,angle=-90}}
\vspace{10mm}

{\bf Fig.5}
\centerline{\hspace{15mm}\psfig{file=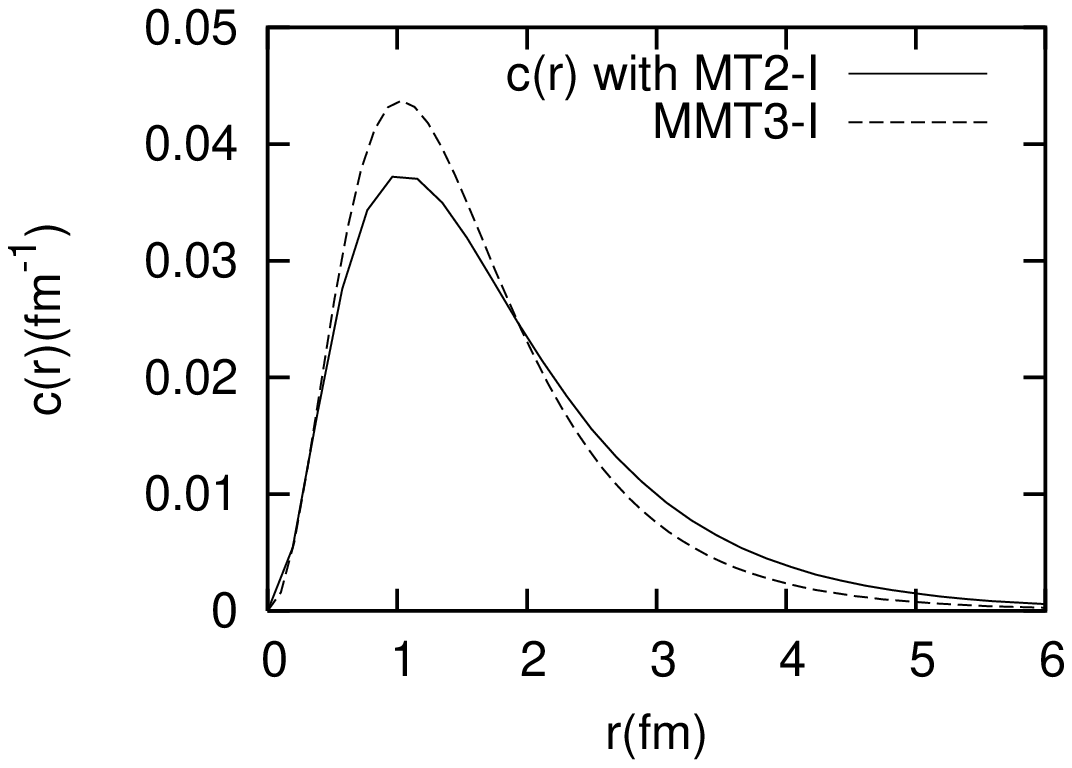,width=120mm,angle=-90}}

\end{document}